# Interactions of Fungi with Concrete: Significant Importance for Bio-Based Self-Healing Concrete

Jing Luo[1], Xiaobo Chen[2], Jada Crump[3], Hui Zhou[2], David G. Davies[4], Guangwen Zhou[2,3], Ning Zhang[1,5*], Congrui Jin[2,3*]

1 Department of Plant Biology, Rutgers University, New Brunswick, NJ 08901, USA

2 Materials Science and Engineering Program, Binghamton University, NY 13902, USA

3 Department of Mechanical Engineering, Binghamton University, NY 13902, USA

4 Department of Biological Sciences, Binghamton University, NY 13902, USA

5 Department of Biochemistry and Microbiology, Rutgers University, New Brunswick, NJ 08901, USA

*Corresponding authors: ningz@rutgers.edu; cjin@binghamton.edu

## Abstract

The goal of this study is to explore a new self-healing concept in which fungi are used as a self-healing agent to promote calcium mineral precipitation to fill the cracks in concrete. An initial screening of different species of fungi has been conducted. Fungal growth medium was overlaid onto cured concrete plate. Mycelial discs were aseptically deposited at the plate center. The results showed that, due to the dissolving of $Ca(OH)_2$ from concrete, the pH of the growth medium increased from its original value of 6.5 to 13.0. Despite the drastic pH increase, *Trichoderma reesei* (ATCC13631) spores germinated into hyphal mycelium and grew equally well with or without concrete. X-ray diffraction (XRD) and scanning electron microscope (SEM) confirmed that the crystals precipitated on the fungal hyphae were composed of calcite. These results indicate that *T. reesei* has great potential to be used in bio-based self-healing concrete for sustainable infrastructure.

## 1. Introduction

Concrete infrastructure suffers from serious deterioration[1,2], and thus self-healing of harmful cracks without high costs or onerous labor have attracted enormous amount of attention. As for how to endow cementitious materials with self-healing properties, many experimental studies and laboratory investigations have been conducted and generated many innovative strategies during the past decades[3-27].

To date, self-healing in concrete has been achieved primarily through three different strategies: autogenous healing, encapsulation of polymeric material, and bacterial production of $CaCO_3$. During the autogenous healing, cracks are filled naturally by means of hydration of unhydrated cement particles and carbonation of dissolved calcium hydroxide as a consequence of exposure to $CO_2$ in the atmosphere[3]. However, this autogenous healing is limited to small cracks (less than 0.2 mm) and requires the presence of water[16]. Encapsulation of polymeric material can fill the cracks in concrete by converting healing agent to foam in the presence of humidity. However, the chemicals released from incorporated hollow fibers behave quite differently from concrete compositions, and they may even cause to further propagate the existing cracks[6].

Due to these drawbacks, the use of the biological repair technique by applying mineral-producing microorganisms becomes highly desirable, as it provides a safe, natural, pollution-free, and sustainable solution to the serious challenge[8-28]. When a calcium source is present, $CaCO_3$, the most suitable filler for concrete due to its high compatibility with cementitious compositions, can be produced through various biomineralization processes. This microbial approach is superior to the other self-healing techniques owing to its excellent microcrack-filling capacity, strong bonding between filler and crack, high compatibility with concrete compositions, favorable thermal expansion, and sustainability[27].

Recent research has demonstrated that some ureolytic bacteria, such as *Bacillus sphaericus* and *B. pasteurii*, have the ability to precipitate calcium carbonate through urea hydrolysis and thus can be used as a powerful tool to heal the cracks[8-12]. However, for each carbonate ion two ammonium ions are produced, leading to excessive nitrogen loading to our environment. To avoid this drawback, metabolic conversion of organic compound to $CaCO_3$ has been proposed by Jonkers et al.[18-20] In this approach, aerobic oxidation of organic acids produces $CO_2$, then leading to the production of $CO_3^{2-}$ in an alkaline environment. Then the presence of a calcium source results in the precipitation of $CaCO_3$. However, this approach requires high concentrations of calcium source[29], which could possibly lead to buidup of high level of salts in concrete. The third pathway to precipitate $CaCO_3$ is known as dissimilatory nitrate reduction[23]. Mineral production is promoted through oxidation of organic compounds through nitrate reduction by means of denitrifying bacteria. However, it has been shown that the efficacy of denitrification approach is much lower than ureolysis regarding the production of $CaCO_3$[30].

**2. Fungi-Mediated Self-Healing Concrete**

While the term "microbe" defines a wide variety of organisms, studies on self-healing concrete are still limited to bacteria[8-27]. Of course, using bacteria has many advantages. For example, bacteria are easy to culture and handle in a laboratory setting and are typically harmless to humans[31]. Moreover, collection and isolation of bacteria are not very complex, as during the years numerous selective media have been introduced for direct isolation of bacteria[32]. On the other hand, however, bacteria do not generally possess sufficient resistance to survive the deleterious environment such as high pH, varied temperature, and dry condition of concrete. So far there has been little success with respect to the long-term healing efficacy and in-depth consolidation, mainly due to the limited survivability and calcinogenic ability of the bacteria. Furthermore, from the economical point of view, the production of bacteria-based self-healing concrete currently results in considerable costs due to the need of aseptic conditions to produce the microbial spores and the use of expensive growth media, making this approach unlikely to be applied in practical applications[33]. In summary, there are still huge challenges to bring an efficient self-healing product to the concrete market with the guaranty that this product can both attain legislative requirements and be cost-effective.

Due to the above-mentioned problems, further investigation on other types of microorganisms having the ability to catalyze calcium mineral precipitation becomes of great potential importance. The overarching goal of the current study is to explore a revolutionary self-healing concept in which bacteria are replaced by fungi to promote calcium mineral precipitation on cracks in concrete infrastructure. Fungi are the most species rich group of eukaryotic organisms after insects with the magnitude of diversity estimated at 1.5M to 3.0M species[34]. Fungi have been investigated mostly due to their important role in organic matter degradation, and their relationship with inorganic matter has mainly been focusing on mineral nutrition via mycorrhizal symbiosis, production of mycogenic organic acids, and lichen bioweathering.

The current study is driven by the following three hypotheses. (1) It is hypothesized that some species of fungi can better adapt to the harsh conditions of concrete including high alkalinity, moisture deficit, and severe oxygen and nutrient limitation[35-50]. (2) It is hypothesized that some species of fungi can promote calcium mineralization in the harsh environment of concrete[51-61]. (3) It is hypothesized that using fungi in biogenic crack repair is more effective than bacteria due to their extraordinary ability to both directly and indirectly promote calcium mineralization[62-73]. The details of the three hypotheses are shown in the Appendix.

To test the hypotheses, this work presents a pilot study to investigate the feasibility of using fungi to promote calcium mineral precipitation to heal cracks in concrete infrastructure. Although many species of fungi have been reported to be able to promote calcium mineralization[71-77], they have never been investigated in the application of self-healing concrete, thus a wide screening of different species of fungi will be conducted.

## 3. Materials and Methods

The following criteria will be used to select the candidates of fungi for self-healing concrete. (1) They should be eco-friendly and nonpathogenic, i.e., pose no risk to human health and are appropriate to be used in concrete infrastructure. Fortunately, fungi that are pathogens are usually pathogenic to plants, and there are comparatively few species that are pathogenic to animals, especially mammals. Among the 100,000 described species of fungi, a little more than 400 are known to cause disease in animals, and far fewer of these species will specifically cause disease in human. Many of the latter are superficial types of diseases that are more of a cosmetic than a health problem. (2) The matrix of young concrete is typically characterized by pH values approximately 13 due to the formation of $Ca(OH)_2$, which is after calcium-silica-hydrate (C-S-H) quantitatively the most important hydration product. Therefore, the fungi placed into the concrete not only have to resist mechanical stresses during the mixing process but also should be able to withstand the high-pH environment for prolonged periods of time. Most promising fungal agents thus should be alkaliphilic spore-forming fungi. The fungal spores, together with nutrients, will be placed into the concrete matrix during the mixing process. When cracking occurs, water and oxygen will find their way in. With enough water and oxygen, the dormant fungal spores will germinate, grow, and precipitate $CaCO_3$ to *in situ* heal the cracks. When the cracks are completely filled and ultimately no more water or oxygen can enter inside, the fungi will again form spores. As the environmental conditions become favorable in later stages, the spores could be wakened again. (3) It is preferred if the genomes of the fungi have been sequenced and are publicly available so that they can be genetically manipulated to enhance their performance in crack repair.

Besides genetically engineered fungi, alkaliphilic fungi could also be found in nature. Through their evolution over millions of years, fungi have developed different strategies to survive and prosper in unfavorable environments. Many species of fungi can grow in alkaline environments where the pH value can often be consistently at about 10[78]. For example, alkaliphilic *Paecillomyces lilacimus* is able to grow well when pH value is between 7.5 and 11.0[78]. *Chrysosporium* spp. isolated from bird nests are also regarded to be alkaliphilic and have a maximum pH for growth at 11[78]. In the current study, the best sources from which alkaliphilic fungi can be isolated will be examined and the field collection will be conducted.

To investigate whether and how fungal hyphae could promote $CaCO_3$ precipitation in concrete, we here will employ material characterization techniques including X-ray diffraction (XRD) and scanning electron microscope (SEM). XRD is a well-established technique to study the structures of mineral crystals, which has been extensively used to identify biominerals at fungi-mineral interfaces[79,80]. SEM has been used for the surface visualization of fungal precipitates[81,82], and in this study SEM will be used to characterize composition and morphology of the solid precipitates.

### 3.1 Isolation of Fungal Strains

The following six species of fungi have been selected and used for this pilot study: *Trichoderma reesei* (ATCC13631), *Aspergillus nidulans* (ATCC38163), *Cadophora interclivum* (BAG4), *Umbeliopsis dimorpha* (PP16-P60), *Acidomelania panicicola* (8D)[83], and *Pseudophialophora magnispora* (CM14-RG38)[84]. *T. reesei* and *A. nidulans* were purchased from the American Type Culture Collection (ATCC). The advantages of using *T. reesei* and *A. nidulans* are their well-understood genetics and a large range of mutants which are affected in a variety of metabolic pathways[85,86].

The other four fungal strains were isolated from the roots of plants that grew in nutrient poor soils. Native roots of pitch pine (*Pinus rigida*), rosette grass (*Dichanthelium acuminatum*), and switchgrass (*Panicum virgatum*) were collected from the New Jersey Pine Barrens in 2011, 2014, and 2016, respectively, and the Sprengel's sedge samples (*Carex sprengelii*) were collected from a subalphine forest in Canadian Rocky Mountains in the province of Alberta, Canada in 2015. The New Jersey Pine Barrens is one of a series of barrens ecosystems along the eastern seaboard. The podzolic soil in this region is sandy, dry, and nutrient poor[87]. However, while lacking nutrients, this habitat supports numerous species that have adapted to the harsh environment. Limited amount of attention has been received on the studies of fungi in this ecosystem, and much remains unrevealed about fungal functions in the pine barrens[87,88]. The soils in the subalphine forest are generally poorly developed, shallow, stony, and have low moisture-holding capacities. The pH values of the soils are approximately 8.1 and moisture availability to plants is very low. In this study, new fungal species uncovered from both the pine barrens and the subalphine forest will be tested in the application of self-healing concrete.

The collected plant root samples were transported on ice to the laboratory for fungal isolation within 24 hours. The plant roots were rinsed by using tap water and then cut into 10-to-20-mm long segments, which were then surface sterilized with 95% ethanol for 30 s, followed by 2 min in 0.6% sodium hypochlorite, 2 min in 70% ethanol, and two final rinses in sterile distilled water. Samples were further cut into 3-mm long small segments, air dried, and placed on 2% malt extract agar (Difco, BD Diagnostic Systems, Sparks, MD, USA) with 0.07 % lactic acid. Lactic acid was used to limit bacterial growth during isolation[89]. Plates were incubated at 25 °C and observed daily in the first two weeks, and then twice a week afterwards for 6 months. Fungal cultures were isolated and purified by subculturing from emergent hyphal tips[83,84]. Spore morphology, if present, and colony characteristics were examined and recorded as morphological data for identification. Fungal cultures have been preserved at -80 °C in glycerol and 4 °C on agar.

### 3.2 Identification of the Isolated Strains

Potato dextrose agar (PDA) that is nutritionally rich in carbohydrates and can stimulate vegetative growth of most fungi was chosen as growth medium. Fungal cultures were grown on PDA (Difco, BD Diagnostic Systems, Sparks, MD, USA) for 7 days. Genomic DNA was extracted from fungal mycelium with the UltraClean Soil DNA Isolation Kit (MoBio Laboratories, Carlsbad, CA, USA). The nuclear ribosomal internal transcribed spacer (ITS) region, the universal fungal barcode marker, was amplified using the primers ITS1 and ITS4[90]. Polymerase Chain Reaction (PCR) was performed with Taq 2X Master Mix (New England BioLabs, Maine, MA, USA). PCR cycling conditions for the ITS consisted of an initial denaturation step at 95 °C for 3 min, 35 cycles of 95 °C for 45 s, 52 °C for 45 s, 72 °C for 1 min and a final extension at 72 °C for 5 min. PCR products were purified with ExoSAP-IT (Affymetrix, Santa Clara, CA, USA) and sequenced with the PCR primers by Genscript, Piscataway, NJ, USA. The fungal stains were identified based on the search results of ITS sequences with BLASTn in GenBank as well as the morphological data.

## 3.3 Preparation of Mortar Specimens and Cement Paste Specimens

Series of mortar specimens were prepared for the survival test of the fungi in the environment of concrete by using Ordinary Portland Cement (CEM I 52.5N), standardized sand (DIN EN 196-1 Norm Sand) and tap water. The water-to-cement weight ratio was 0.5 and the sand-to-cement weight ratio was 3. The specimens were made according to the standard procedure NBN EN 196-1[91]. They were then poured into 60 mm Petri dishes (9 ml per dish) and cured at 100% relative humidity and 22 °C for 7 days.

Cement paste specimens were prepared to investigate the pore size distribution of aging specimens. Ordinary Portland Cement (CEM I 52.5N) was mixed with tap water in a water-to-cement weight ratio of 0.5. Liquid paste was poured in molds with dimensions of 40 mm × 40 mm × 40 mm and cured at 100% relative humidity and 22 °C for 1, 3, 5, 7, 14, or 28 days.

Air-entrained cement paste specimens were prepared to investigate the effect of air-entraining on the pore size distribution of the specimens. Ordinary Portland Cement (CEM I 52.5N) was mixed with tap water containing the air-entraining agent in a water-to-cement weight ratio of 0.5. Eucon AEA-92 (Euclid Chemical, Cleveland, OH, USA) was dosed at a rate of 100mL to 260 mL per 100 kg of the total cementitious material. Liquid paste was poured in molds with dimensions of 40 mm × 40 mm × 40 mm and cured at 100% relative humidity and 22 °C for 28 days.

## 3.4 Survival Test of Fungi in the Environment of Concrete

To check the effect of the highly alkaline environment of concrete on the fungal growth behavior, growth medium was prepared using PDA (Difco, BD Diagnostic Systems, Sparks, MD, USA) with or without the addition of the inert pH buffer 3-(N-morpholino)propanesulfonic acid (MOPS, 20mM, pH 7.0) (Fisher Scientific, Pittsburgh, PA, USA). 10 ml growth medium was overlaid onto each cured concrete plate. A mycelial disc with a diameter of 5 mm of each fungal strain was removed from 7-day-old cultures using a cork borer, and was aseptically deposited at the center of each 60 mm Petri dish containing growth medium with or without concrete. Sterile PDA plugs were used as the negative inoculum control. After inoculation, the Petri dishes were incubated in natural daylight conditions at 25 °C and 30 °C, respectively, for three weeks. Radial growth measurements were recorded along two perpendicular diameters. The fungal growth was also evaluated via optical microscopy. The fungal samples were prepared by the tape touch method[92] and observed with an optical microscope (Carl Zeiss model III, Zeiss, Jena, Germany).

For each type of fungal strain, totally eight different types of plates were tested in this study, which were abbreviated as follows: PDA incubated at 30 °C (PDA30), PDA incubated at 25 °C (PDA25), PDA with MOPS incubated at 30 ºC (MPDA30), PDA with MOPS incubated at 25 ºC (MPDA25), PDA with concrete incubated at 30 ºC (CPDA30), PDA with concrete incubated at 25 ºC (CPDA25), PDA with both concrete and MOPS incubated at 30 ºC (CMPDA30), and PDA with both concrete and MOPS incubated at 25 ºC (CMPDA25). All the tests were done independently in triplicates.

### 3.5 pH Measurements

Since PDA is a gelatinous substance, we could measure the pH of each plate. The pH of each plate was measured by taking five independent measurements using an Orion double junction pH electrode (Thermo Fisher Scientific, Waltham, MA, USA). pH measurements were recorded after the plates were incubated for three weeks.

### 3.6 Microscopic Characterization of Biominerals Produced by Fungi

#### 3.6.1 X-Ray Diffraction (XRD)

The solid precipitates associated with fungal hyphae (from the *T. reesei* cases of CPDA30 and CMPDA30) were identified by XRD analysis. A Siemens-Bruker D5000 powder diffractometer with Cu-Kα radiation in the theta/theta configuration was used. The diffractometer was operated at 40 kV and 30 mA. Measurements were made from 10° to 80° 2θ at a rate of 1°/min with a step size of 0.02° 2θ. Isolation of the fungal precipitates was performed by dissolving the fungi in NaOCl and repeated washings in methanol according to the published protocol[80].

#### 3.6.2 Scanning Electron Microscope (SEM)

The solid precipitates associated with fungal hyphae (from the *T. reesei* cases of CPDA30 and CMPDA30) and the NaOCl-isolated crystals were analyzed using a Zeiss Supra 55 VP Field Emission SEM with an EDAX Genesis energy-dispersive X-ray spectrometer (EDS) at accelerating voltages of 5 kV to 20 kV. The fungal samples were completely dried in the oven at 50 ºC for 2 days, and then were mounted on aluminum stubs and sputter-coated with carbon to ensure electrical conductivity for the examination of crystal morphology and distribution. The elemental composition of the precipitates was investigated by EDS analysis.

### 3.7 Pore Size Distribution of Cement Paste Specimens

To determine the pore size distribution in aged cement paste specimens, the mercury intrusion porosimetry (MIP) method was used to measure the matrix pore sizes in 1, 3, 5, 7, 14, and 28 days cured cement paste specimens using a Model AMP-30K-A-1 (Porous Materials, Ithaca, NY, USA). Aged cement paste specimens were cut to smaller cubes of 4 mm × 4 mm × 4 mm, which were subjected to cryo-vacuum evaporation for two weeks for pore water removal before MIP tests. MIP tests were conducted according to the published protocol[93].

## 4. Results and Discussion

### 4.1 Identification of BAG4, PP16-P60, 8D, and CM14-RG38

Strain PP16_P60 was isolated from the pitch pine in the Pygmy Pine Plains of the New Jersey Pine Barrens. It has 100% ITS sequence similarity to *Umbeliopsis dimorpha* ex-type culture CBS110039 (NR_111664), and thus it is identified as *Umbeliopsis dimorpha*. BAG4 was recovered from the Sprengel's sedge samples (*Carex sprengelii*) collected from a subalphine forest in Canadian Rocky Mountains in the province of Alberta, Canada. The blast result indicates its phylogenetic position in the genus *Cadophora*, and it is further identified as *Cadophora interclivum* based on multigene and morphological analyses. 8D is associated with switchgrass roots in the New Jersey Pine Barrens. It has 92% or less ITS sequence (KF874619) similarities to any known or described species with accessible ITS sequences in GenBank, such as *Mollisia fusca* CBS486.48 (AY259137) and *Loramyces macrosporus* AFTOL-ID 913 (DQ471005), and further identified as *Acidomelania panicicola*. CM14_RG38 was found from the rosette grass in Collier Mills in the New Jersey Pine Barrens. Its ITS sequence (KP769835) has 96% or less similarities to other *Pseudophialophora* species, such as *Pseudophilophora eragrostis* CM12m9 (KF689648) and *Pseudophilophora whartonensis* WSF14RG66 (KP769834), and it is identified as *Pseudophialophora magnispora*[84].

**4.2 Fungal Growth in the Environment of Concrete**

The fungal growth in each type of plate has been shown in Fig. 1. Optical microscopic analysis of each case has been shown in Fig. 2. The growth rates of all the tested species are showed in Table 1. The pH measurement results, as listed in Table 2, have shown that, due to the dissolving of $Ca(OH)_2$ from concrete, the pH of the growth medium in the cases of CPDA30 and CPDA25 increased from 6.5 to 13.0. Only *T. reesei* (ATCC13631) has been found to be able to grow well on the concrete plates. At 30 °C, its growth rates reached 2.6 mm/day in the cases of both CPDA30 and CMPDA30. Abundant conidia were observed from the concrete plates and had similar morphology compared to those produced on the plates without concrete, i.e., the cases of PDA30, PDA25, MPDA30, and MPDA25. However, no growth of *T. reesei* was found on any concrete plates at 25 °C, i.e., the cases of CPDA25 and CMPDA25. The other five species had no growth on any concrete plates, although they grew on most of the non-concrete plates. The MOPS buffer significantly decreased the fungal growth, which is probably due to the relatively high concentration used in the experiments. Agar plug controls without any inoculum showed no fungal growth.

*Table 1. Average growth rates (mm/day) of the six fungi species on PDA, MPDA, CPDA, and CMPDA at 25 °C and 30 °C, respectively, at day 21 after inoculation (n = 6).*

|  | PDA30 | PDA25 | MPDA30 | MPDA25 | CPDA30 | CPDA25 | CMPDA30 | CMPDA25 |
|---|---|---|---|---|---|---|---|---|
| *Trichoderma reesei* (ATCC13631) | 2.6 a | 2.6 a | 1.0 b | 0.8 b | 2.6 a | 0 | 2.6 a | 0 |
| *Aspergillus nidulans* (ATCC38163) | 2.6 a | 2.6 a | 0.5 b | 0.7 b | 0 | 0 | 0 | 0 |
| *Cadophora interclivum* (BAG4) | 0.6 b | 2.1 a | 0 | 0.6 b | 0 | 0 | 0 | 0 |
| *Umbeliopsis dimorpha* (PP16-P60) | 2.6 a | 2.6 a | 1.0 b | 1.0 b | 0 | 0 | 0 | 0 |
| *Acidomelania panicicola* (8D) | 2.1 a | 1.9 a | 0.9 b | 0.9 b | 0 | 0 | 0 | 0 |
| *Pseudophialophora magnispora* (CM14-RG38) | 2.6 a | 2.1 a | 0 | 0 | 0 | 0 | 0 | 0 |

*Table 2. pH measurement results of the six fungi species on PDA, MPDA, CPDA, and CMPDA at day 21 after inoculation.*

|  | PDA | MPDA | CPDA | CMPDA |
|---|---|---|---|---|
| Control | 5.1 | 6.8 | 13.1 | 11.5 |
| *Trichoderma reesei* (ATCC13631) | 6.5 | 7.2 | 13.0 | 11.9 |
| *Aspergillus nidulans* (ATCC38163) | 6.8 | 7.1 | 12.2 | 10.9 |
| *Cadophora interclivum* (BAG4) | 6.3 | 7.1 | 12.0 | 11.4 |
| *Umbeliopsis dimorpha* (PP16-P60) | 6.1 | 7.1 | 12.1 | 11.3 |
| *Acidomelania panicicola* (8D) | 6.8 | 7.1 | 12.6 | 11.9 |
| *Pseudophialophora magnispora* (CM14-RG38) | 6.9 | 7.0 | 12.0 | 11.0 |

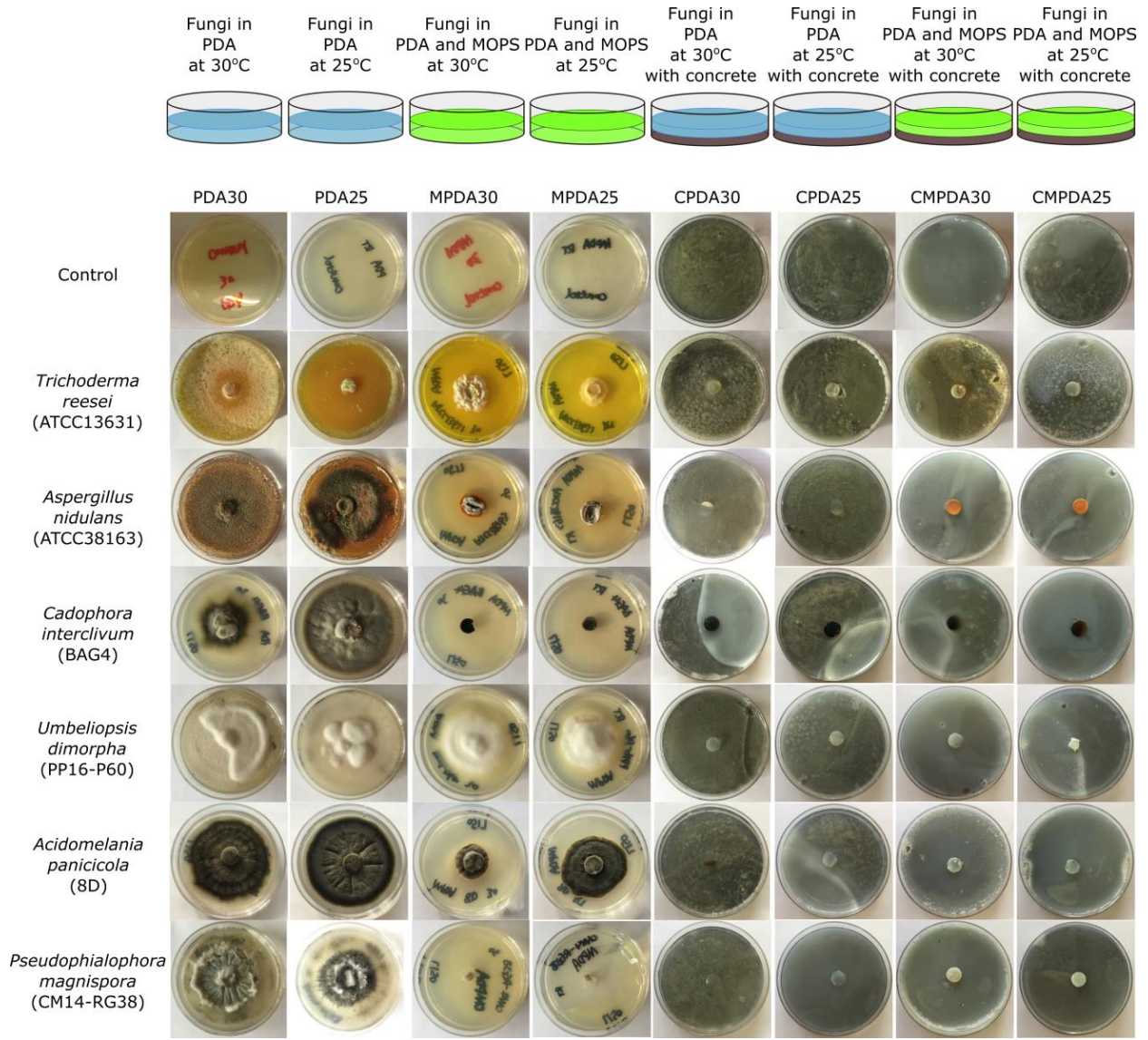

*Figure 1. T. reesei spores germinated on concrete into hyphal mycelium and grew equally well with or without concrete. In comparison, the other five species did not grow on concrete.*

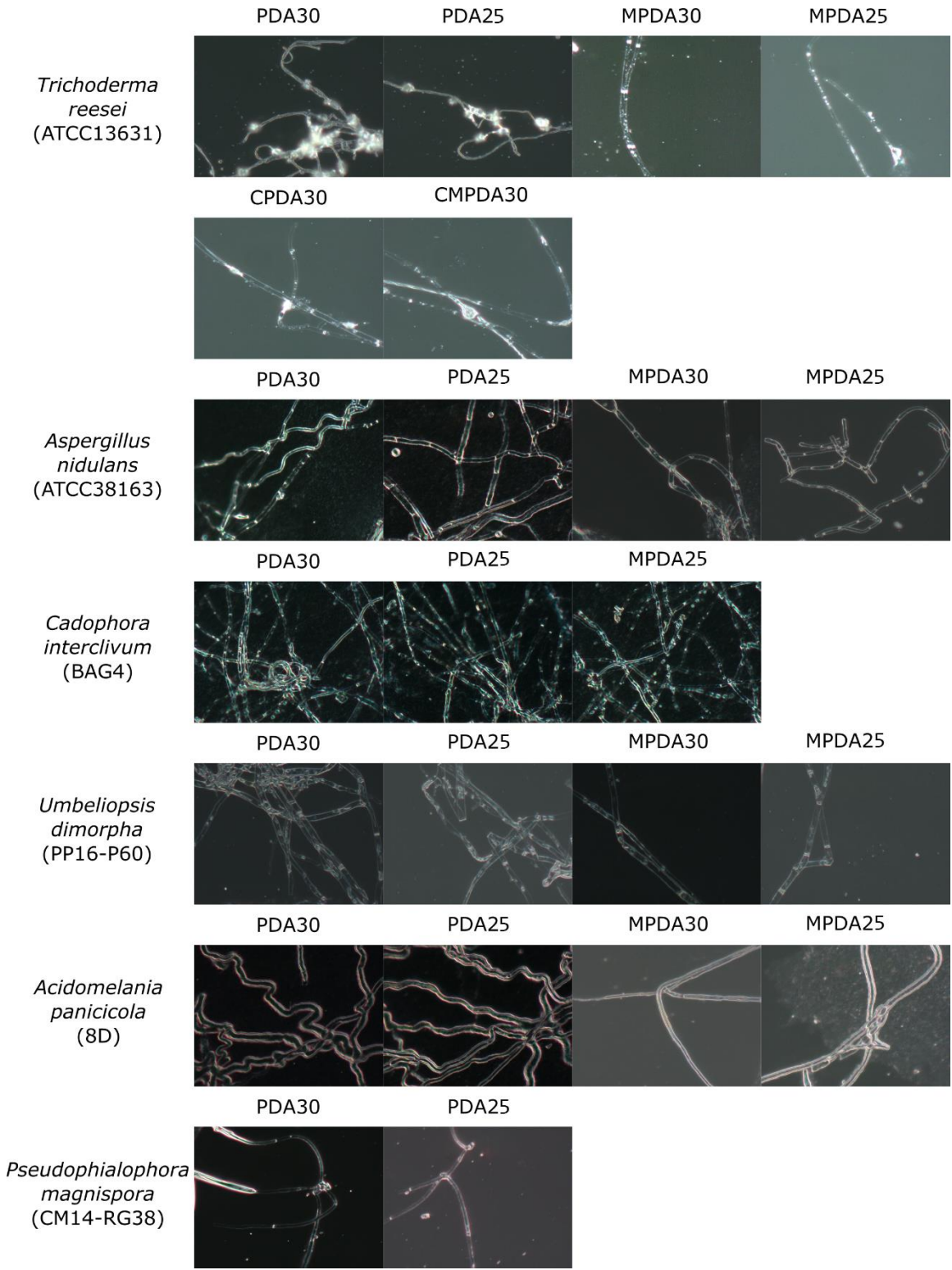

*Figure 2. Microphotographs of optical microscopy (1000X, Carl Zeiss model III) showing that T. reesei spores germinated on concrete into hyphal mycelium and grew equally well with or without concrete.*

**4.3 Identification and Morphology of the Fungal Precipitates**

The results from XRD analysis are shown in Fig. 3. The data strongly suggested that *T. reesei* hyphae can promote calcium carbonate precipitation. For the precipitates associated with the fungal hyphae, the sharp peak at around 30º 2θ suggests the presence of highly crystalline phases of the calcium carbonate mineral calcite. The mortar specimens obtained from the parallel experiment performed with the agar control without fungi were mainly composed of highly crystalline phases of quartz and calcite. The carbonation is the result of the dissolution of $CO_2$ in the concrete pore fluid and its reaction with $Ca(OH)_2$.

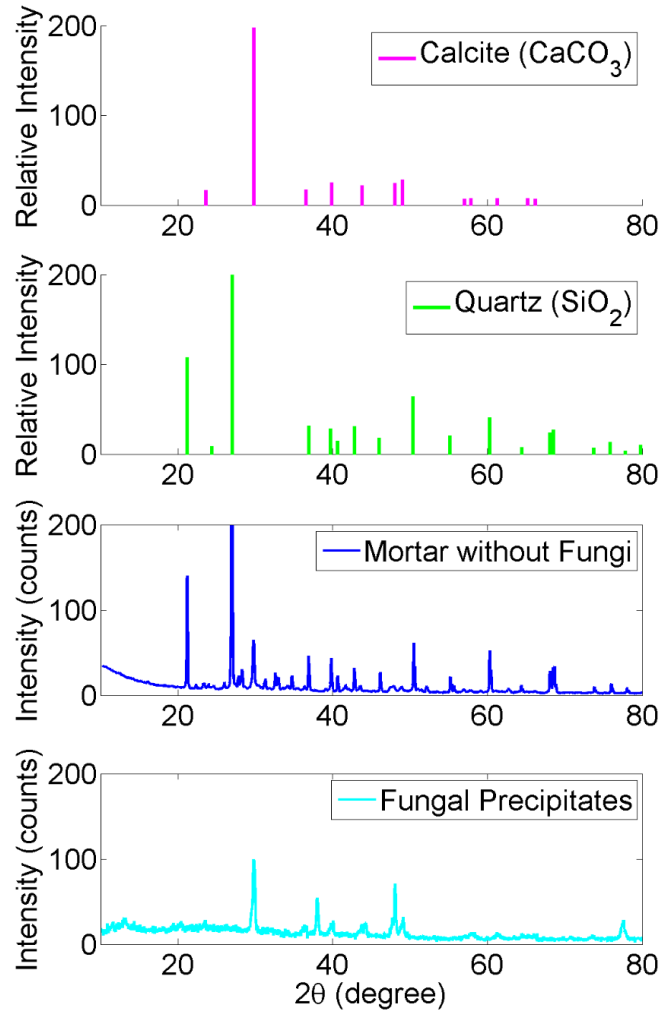

*Figure 3. XRD results for crystalline precipitates collected from mortar specimens cured with and without T. reesei hyphae. For comparison, reference diffractograms of quartz ($SiO_2$) and calcite ($CaCO_3$) mineral standards from the International Centre for Diffraction Data (ICDD) are included.*

The SEM images are shown in Fig. 4. It can be seen that a large amount of mineral crystals grew in the *T. reesei*-inoculated medium. The mineral crystals showed evidence of fungal involvement. Wire-shaped traces having an average thickness of 2 µm to 3 µm were found on the surface of the minerals, which presumably occurred in the space occupied by the fungi. These traces also indicated that fungal hyphae served as nucleation sites during the mineral precipitation process. EDS analysis demonstrated that the crystal is composed of Ca, C, and O with an atomic percentage closely matching that of $CaCO_3$, implying that the crystal is composed solely of calcium carbonate. In sharp contrast to the fungi-inoculated medium, the amount of formed crystals in the fungi-free control medium was much less. Furthermore, in the control medium, no sign of fungi involvement was observed during the mineral precipitation.

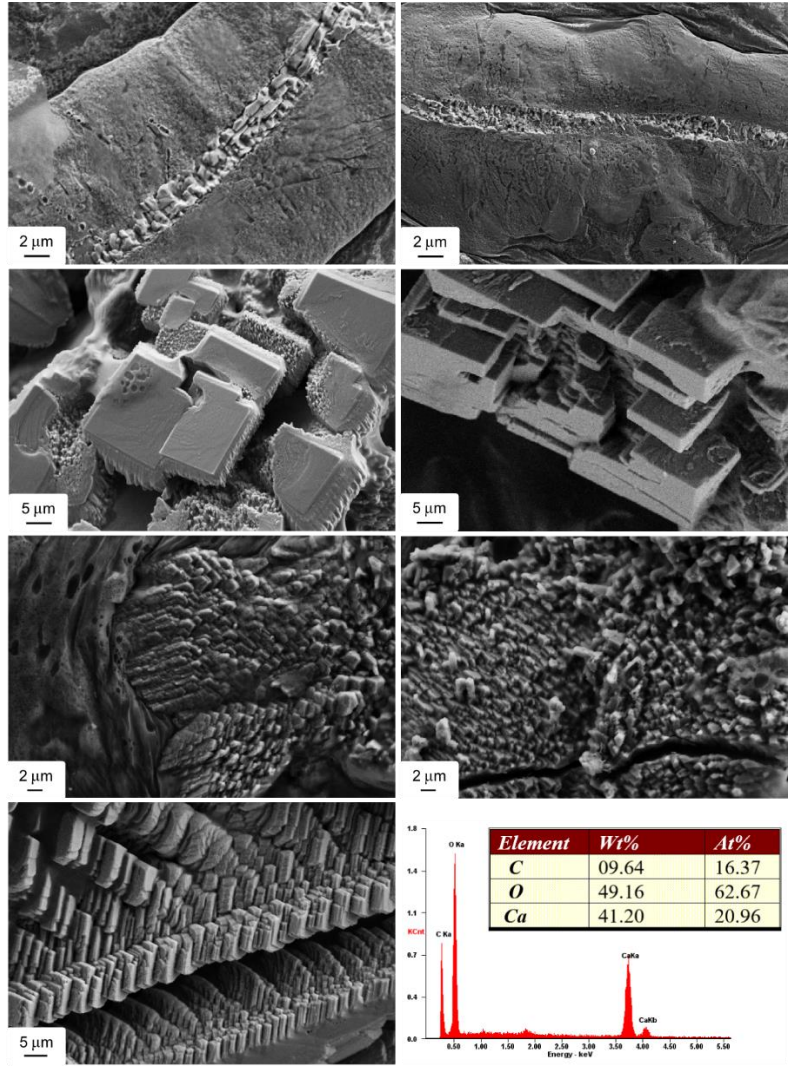

*Figure 4. SEM and EDS spectra of the calcium carbonate precipitation in the T. reesei-inoculated medium.*

**4.4 Discussion on Embedment of Healing Agent in Concrete Matrix**

In this section, how to embed the healing agents, i.e., fungi spores and nutrients, into concrete will be briefly discussed. If the typical fungi spore is larger than the pore sizes in concrete, when the healing agents are directly put into cement paste specimens, the majority of spores will be squeezed and crushed due to the pore shrinkage during the hydration process, leading to loss of viability and decreased mineral-forming capacity. As we measured by using the mercury intrusion porosimetry (MIP) method, the matrix pore diameter sizes in 28 days cured specimens decreased to less than 0.1 μm, as shown in Fig. 5(a), which cannot accommodate fungal spores with typical diameters larger than 3 μm, as shown in Fig. 6. Therefore, encapsulation or immobilization of fungi spores in a protective carrier becomes essential.

In several previous studies of bacteria-based self-healing concrete, bacteria spores were placed inside discrete tubular or spherical capsules to increase the viability of bacteria, since capsules help to resist mechanical forces during the concrete preparation. When damage happens, the capsules rupture and the self-healing process commences through release of the healing agent. However, much lower mechanical strength was noted in capsules-incorporated concrete specimens, mainly attributed to the empty spaces after capsule activation[25]. Therefore, immobilization of bacteria spores into silica gel, polyurethane, and hydrogel were used to address the encapsulation drawbacks[17,26]. However, these materials are either expensive or inappropriate to be used for concrete.

We propose that air-entraining agents could be utilized to create extra air voids in concrete matrix to facilitate the housing of the fungal spores. The matrix pore diameter sizes in 28 days cured air-entrained specimens are shown in Fig. 5(b). It can be seen that the amount of entrained air voids increases with increasing amount of air-entrained agents.

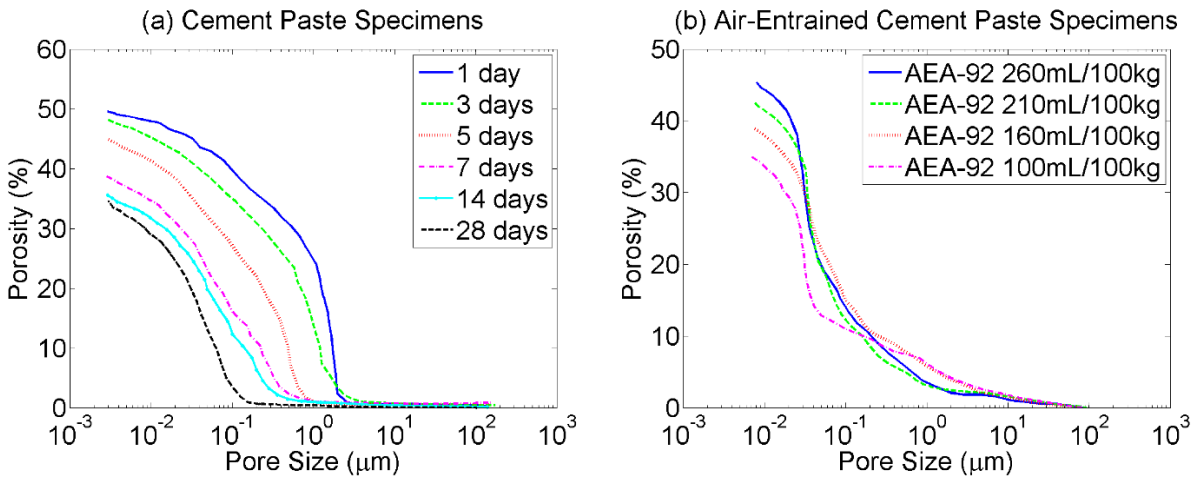

*Figure 5. (a) Pore size of cement paste specimens with different curing time prepared with a water-to-cement weight ratio of 0.5 measured by MIP tests. (b) Effect of the amount of air-entraining agent on pore size of cement paste specimens prepared with a water-to-cement weight ratio of 0.5 cured for 28 days. MIP tests were conducted using a Model AMP-30K-A-1(Porous Materials, Ithaca, NY, USA).*

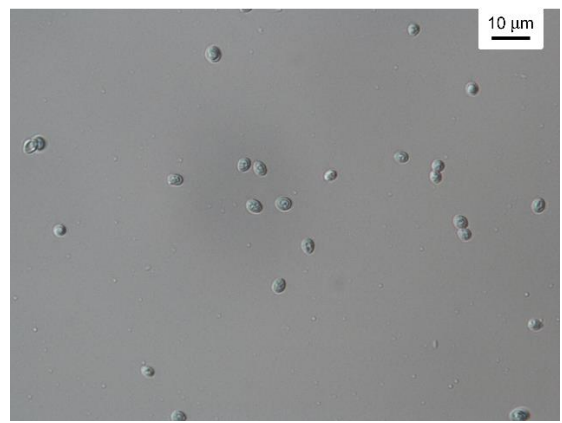

*Figure 6. The diameter of T. reesei spores (round to oval in shape) used in the experiments appeared to be typically in the range of 3.5 μm to 4.5 μm.*

## 5. Concluding Remarks and Future Work

In the current study, a new self-healing concept has been explored, in which fungi were used to promote calcium mineral precipitation to heal cracks in concrete infrastructure. An initial screening of different species of fungi has been conducted. The experimental results showed that, due to the dissolving of Ca(OH)$_2$ from concrete, the pH of the growth medium increased from its original value of 6.5 to 13.0. Despite the drastic pH increase, the microscopic analysis showed that *T. reesei* (ATCC13631) spores germinated into hyphal mycelium and grew equally well with or without concrete. In comparison, *A. nidulans* (ATCC38163), *C. interclivum* (BAG4), *U. dimorpha* (PP16-P60), *A. panicicola* (8D), and *P. magnispora* (CM14-RG38) did not grow on concrete. We employed material characterization techniques including XRD and SEM, both of which confirmed that the crystals precipitated on the fungal hyphae were composed of calcite.

In addition, there is no evidence in the scientific literature indicating that *T. reesei* is a human pathogen. Despite its widespread presence in tropical soils, there are no reports of the species causing adverse effects in aquatic or terrestrial plants or animals in the tropics. In the pathogenicity study of *T. reesei* made with mice, guinea-pigs, and rabbits[94], the results also confirmed that *T. reesei* can be regarded as non-pathogenic and it is unlikely to cause infection in healthy or debilitated humans. *T. reesei* is susceptible to major clinical antifungal drugs that could be used for treatment in the unlikely event of infection. In fact, it has a long history of safe use in industrial-scale production of carbohydrase enzymes, such as cellulase, due to its capacity to secrete large amounts of cellulolytic enzymes[95]. Repeated exposure to commercial enzyme preparations produced by *T. reesei* rarely causes allergic reactions in humans. Therefore, we could conclude that *T. reesei* has great potential to be safely used in bio-based self-healing concrete for sustainable infrastructure. Of course, a thorough assessment should be conducted to investigate the possible immediate and/or long-term harmful effects on environment and/or human health prior to its use as healing agents in concrete infrastructure.

As future work, the effects of the various factors influencing fungal survival in the harsh environment of concrete and/or fungal calcium precipitation will be investigated, including temperature, growth medium composition, fungal spore concentration, and different chemical admixtures often added to concrete for the purpose of water-reducing, set-retarding, set-accelerating, corrosion inhibition, shrinkage reduction, and workability enhancement, etc. In addition, further studies to elucidate the effect of incorporating fungi-based healing agents on concrete properties such as strength, permeability and carbonation resistance are needed but are beyond the present study. The interface between fungal precipitates and concrete crack edge will also be microscopically characterized, and its bond coherence will be studied by atomic force microscope (AFM), which is an important criterion that should be considered to avoid new crack formation.

This research will also benefit many other applications as well, such as metal remediation, carbon sequestration, and enhanced oil recovery. (1) The discharge of heavy metals from mining and metal-processing industries has resulted in serious contamination in both soil and groundwater. Conventional methods, such as ion exchange method and chemical reaction method, are either ineffective or require high investments[96]. In comparison, microbial CaCO$_3$-based coprecipitation offers a low-cost and environment-friendly solution[97]. This study will be helpful in providing insight on how fungi could effectively bind metal ions. (2) In recent years, $CO_2$ sequestration has become an attractive choice to mitigate $CO_2$ emission[98]. Compared with geological sequestration, which may suffer from upward leakage of $CO_2$ through fractures[99], microbial $CO_2$ fixation offers a reliable, low-cost, and economically sustainable storage strategy[100]. The current research will advance our understanding of how fungi could be used in such applications. (3) Enhanced oil recovery has become essential to improve the overall extraction percentage of crude oil[101], but the thermal or chemical flooding processes associated with enhanced oil recovery are environmentally hazardous and extremely expensive[101]. In comparison, microbial enhanced oil recovery, in which microbes are used to plug high-permeability zones for a redirection of waterflood, thus improving the yield of reservoir oil, offers a low-cost and environment-friendly strategy[102]. The fundamental knowledge advanced by the current study is essential to understand the role of fungi in enhanced oil recovery.

**Appendix: Three Hypotheses on Fungi-Mediated Self-Healing Concrete**

The current study is driven by the following three hypotheses.

(1) It is hypothesized that some species of fungi are able to better adapt to the harsh conditions of concrete including high alkalinity, moisture deficit, and severe oxygen and nutrient limitation. Fungi are well-known for their remarkable ability to survive extreme environments such as those characterized by limited nutrient availability, extreme temperatures and pressures, high salinity, high radiation, intense ultraviolet light, and variable acidity[35-48]. For example, some can be found in the Arctic and Antarctic cold deserts[36], the Sahara Desert in Africa[37], the hypersaline Dead Sea[38], as well as the deep-sea sediments of the Indian Ocean at depths of about 5000 m[39]. Some can even survive the intense ultraviolet light and cosmic radiation during space travel[40]. Most relevantly, rocks are often considered as an extreme environment for fungi due to nutrient deprivation, exposure to ultraviolet light, and low humidity, but fungi were reported to be able to survive in a wide variety of substrates, such as limestone, marble, sandstone, granite, and gypsum, where they make up a critical component of both epilithic and endolithic communities of microbes[41-48]. The ability of most fungi to form hard spores that can survive almost any natural environment is another important fungal adaptation[49]. When exposed to environmental stresses such as limited nutrients, they produce dormant and highly resistant cells termed spores that can survive in this dormant state for long periods of time, waiting for more favorable conditions. Faced with the challenge of surviving prolonged periods of dormancy, spores have evolved various mechanisms to survive environmental assaults that would normally kill the fungi, and as a result, these spores can endure many years of hardship. For example, in 2004, spores from a fungus that lived roughly 400,000 years ago were germinated in a laboratory in India[50].

(2) It is hypothesized that some species of fungi are able to promote calcium mineralization in the harsh environment of concrete. According to Verrecchia[51], plenty of near-surface limestones, calcic and petrocalcic horizons in soils were secondarily cemented and indurated with calcite, whewellite, and weddellite. Although, in part, these phenomena can be ascribed to physico-chemical processes, the existence of calcified fungal hyphae in calcareous soils and limestone in a variety of localities implies that fungi could play a significant role in secondary $CaCO_3$ precipitation[52-57]. It has been known for a long time that oxalate salts, especially whewellite and weddellite, are often found with fungal filaments in soils, leaf litter, and lichen thalli[58,59]. According to Verrecchia et al.[60], oxalate could be degraded to carbonate, particularly in a semi-arid environment, in which such a process may act in the cementation of limestones[60].

The formation of calcium minerals by *Serpula himantioides* and a limestone fungal isolate identified as a *Cephalotrichum* sp. has been studied by Burford et al.[61]. X-ray diffraction of crystalline precipitates on the *S. himantioides* hyphae showed that they were composed of calcite and some whewellite, and the crystalline precipitates on the hyphae of the limestone isolate were composed solely of calcite or of a mixture of calcite and weddellite. This study provided direct experimental evidence for the precipitation of calcite and secondary calcium minerals on fungal hyphae in a low-nutrient environment.

(3) It is hypothesized that using fungi in biogenic crack repair is more effective than bacteria due to their extraordinary ability to both directly and indirectly promote calcium mineralization. It is widely believed that filamentous fungi possess distinctive advantages over other microbial groups to be used as biosorbent materials to attract and hold metal ions because of their superior wall-binding capacity and extraordinary metal-uptake capability[62-65]. Although the peculiar mechanisms leading to calcium mineralization by fungi remains incompletely understood, but it is widely believed that there are several different processes involved in the calcification of fungal hyphae, such as the cation binding onto fungal cell walls, and the formation of calcite through fungal excretion of hydrogen ions or organic acids[66-68].

Cation binding by fungi is a metabolism-independent process of binding ions onto cell walls and other external surfaces, resulting in mineral nucleation and deposition[66,67]. Bound calcium cations often interact with soluble $CO_3^{2-}$, leading to $CaCO_3$ deposition on the fungal filaments. Calcite formed in the aqueous phase could also nucleate onto the hyphae. Since the biosorption of metal ions onto fungal cell walls is a metabolism-independent process, dead and metabolically inactive fungal hyphae can also act as nucleation sites of further calcium carbonate precipitation[68]. An important attribute that places fungi in a different kingdom from bacteria is the chitin in their cell walls, which is a modified polysaccharide that contains nitrogen[69]. According to Manoli et al., chitin is a substrate that significantly lowers the required activation energy barrier for nucleus formation so that calcite can readily nucleate and subsequently grow on it[70]. In other words, the interfacial energy between the fungi and the mineral crystal becomes much lower than that between the mineral crystal and the solution.

On the other hand, the excretion of organic acids, especially oxalic acid, by fungal filaments plays an important role in the re-precipitation of secondary calcium minerals in the $CaCO_3$-rich environments[51]. Oxalic acid produced by *Aspergillus niger* has been demonstrated to be able to react with calcium ions and $CaCO_3$ to form calcium oxalate by Sayer et al.[71]. *A. niger* and *S. himantioides* were demonstrated to be able to precipitate calcium oxalate when cultured on gypsum by Gharieb et al.[72]. The fungal excretion of oxalic acid and the precipitation of calcium oxalate may result in the dissolution of the internal pore walls of the limestone matrix, making the solution enriched in $CO_3^{2-}$. As the solution passes through the pore walls, $CaCO_3$ could recrystallize as a consequence of the decreased level of $CO_2$[60]. Biodegradation of oxalate by means of microbial activity can also result in transformation into $CO_3^{2-}$, leading to $CaCO_3$ precipitation in the pore interior[61]. During decomposition of fungal filaments, $CaCO_3$ crystals could be released to work as further secondary calcite precipitation sites[51,60]. The $CO_2$ production results from both oxalate oxidation and fungal respiration can cause $CO_3^{2-}$ concentration in the local environment and thus favor more $CaCO_3$ precipitation. According to Verrecchia et al., a large amount of the secondary $CaCO_3$ found in soils and surficial sediments originates by such a process[73].

**Acknowledgement**

Congrui Jin and David G. Davies were funded by the Research Foundation for the State University of New York through the Sustainable Community Transdisciplinary Area of Excellence Program (TAE-16083068). Congrui Jin also thanks the support from the Small Scale Systems Integration and Packaging (S3IP) Center of Excellence, funded by New York Empire State Development's Division of Science, Technology and Innovation. Jada Crump, affiliated with Westchester Community College, Valhalla, NY, USA, conducted her research in the Department of Mechanical Engineering at Binghamton University in the summer of 2017 when she was a visiting research assistant in Congrui Jin's research group. Her visit was supported by the State University of New York Louis Stokes Alliance for Minority Participation Program (SUNY LSAMP). The authors gratefully thank the anonymous reviewers for their critical comments.